\documentclass[%
reprint,
longbibliography,
superscriptaddress,
 amsmath,amssymb,
 aps,
 prb,
]{revtex4-1}

\usepackage{graphicx}
\usepackage{dcolumn}
\usepackage{bm}
\usepackage{tabularx}

\begin{document}

\preprint{APS/123-QED}

\title{Strong local variational approach for superconductivity theory, and the principles of coherent interaction and action- counteraction}

\author{ChaoFan Yu}
\affiliation{School of Physics and Information Engineering, Guangdong University of Education, Guangzhou 510303, China}

\author{Xuyang Chen}
\email{2022010138@m.scnu.edu.cn}
\affiliation {Key Laboratory of Atomic and Subatomic Structure and Quantum Control (Ministry of Education), Guangdong Basic Research Center of Excellence for Structure and Fundamental Interactions of Matter, School of Physics, South China Normal University, Guangzhou 510006, China}

\author{ZhiHua Luo}
\affiliation{School of Physics and Information Engineering, Guangdong University of Education, Guangzhou 510303, China}

\begin{abstract}
For the two-mode electron pairing, we put forward a local stacking force pairing mechanism caused by the strong local fluctuation principle, with two straight pairing orbits, in there the tying Cooper pairing $C_{-k\downarrow}C_{k\uparrow}e^{ik\cdot r}$ replaces the itinerant Cooper pairing $C_{-k\downarrow}C_{k\uparrow}$. Based on the principles of coherent interaction and action-counteraction, the strong local variational theory has been constructed, in which the energy extremum equation and the energy gap equation form a pair of self-consistent equations, with the local variational parameter $\lambda$, the energy gap $\Delta$, and the energy integral upper limit $\hbar \omega_{0}$. Since the energy extremum equation and the energy gap equation are restricted each other, when $\hbar \omega_{0}(j)$ arrives at the energy integral upper limit (i.e. the cut-off value), $\lambda$ and $\Delta$ are also tended to their certain values at the same time. As the results of the strong local variational theory, the coupling strength $Vg(0)$ reduces to $\tilde{V}g(0)=e^{-\left(1-\alpha_{1}\right)^{2} k^{2} / 4 \lambda^{2}} Vg(0)$, while the Cooper pair reduces to $\widetilde{C_{k \uparrow} C_{-k \downarrow}}=e^{-\left(1-\alpha_{1}\right)^{2} k^{2} / 4 \lambda^{2}} C_{k \uparrow} C_{-k \downarrow}$. For the weak coupling, $\alpha_{1}=1$, when $V g(0)=0.1, \Delta_{\mathrm{A \cdot C}}=108 \Delta_{\text{BCS}}$, but $Vg(0)=$ $0.2, \Delta_{\mathrm{A \cdot C}}=28 \Delta_{\text{BCS}}$. For the strong coupling, $\alpha_{1}=0$, if $Vg(0)=1.4$ , $\tilde{V} g(0)$ will reduce to 0.2 and $\widetilde{C_{k \uparrow} C_{-k \downarrow}}=0.14 C_{k \uparrow} C_{-k \downarrow}$ (the smaller Cooper pair) from beginning to end with the $c_{k}$-electron energy $\epsilon_{k}$ since $k^{2} / \lambda^{2}=$ const. On the other hand, we obtain that $\Delta_{\mathrm{A \cdot C}}=$ $0.5676 ~\text{eV} \gg \hbar \omega_{\text{D}} \gg \Delta_{\text{BCS}}$, and the strong local fluctuation stacking force $\widetilde{V}_{\text{st}}=0.264 ~\text{eV}  \gg \hbar \omega_{\text{D}}(0.01 ~\text{eV} )$. In particular, since $k^{2} / \lambda^{2}=$ const, the energy integral upper limit $\hbar \omega_{0}$ and the energy gap $\Delta$ will be given as the unique solution. Importantly, since $K^2\uparrow \rightarrow \lambda^2 \uparrow \rightarrow \tilde{V}_{\text{st}}\uparrow \rightarrow K^2\uparrow \rightarrow \lambda^2 \uparrow \rightarrow \tilde{V}_{\text{st}}\uparrow \cdots$, the development of $\tilde{V}_{st}$ has no upper limit, as this reason, $\Delta=\tilde{V}_{st}\sum_k \langle C_{-k}C_k\rangle$ has also no upper limit. Nothing that the cut-off frequency $\hbar \omega_0(c)$ (the energy integral upper limit $\epsilon_k(c)$) increases as the coupling strength $\tilde{V}_{st}g(0)$ increases, with $\tilde{V}_{st}g(0) \gg 1.0$, we will see $\epsilon_{k}(c) \gg \hbar \omega_{D}$, then correspondingly we have $\tilde{V}_{st} \gg \hbar \omega_{D}$ and $\Delta \gg \hbar \omega_{D}$, this is the dream that the ambient temperature superconductivity could be realized.
\end{abstract}

\maketitle

\section{Introduction}
Superconductivity, a remarkable macroscopic quantum phenomenon, touches unexpected and different parts of physics and science. A milestone discovery by Bednorz and Mller \cite{1} of high $T_{\text{c}}$ superconductivity in cuprates, in1986, suggested that the dream of ambient temperature could be realized. The resonating valence bond theory of high $T_{\text{c}}$ superconductivity that began as a response to understand cuprate superconductivity by Anderson\cite{2} and collaborators\cite{3,4} gave new hope and created a fertile ground for high $T_{\text{c}}$ superconductivity\cite{5,6,7,8}. Cuprates cannot be described by the Fermi liquid model. One such theory uses the idea of resonating valence bonds, mainly that the electron from neighboring atoms can form valence bonds to form pairs that can be doped to form a stable superconducting ground state. Another possibility is that fluctuations of antiferromagnetism glue two electrons together to form Cooper pairs \cite{9,10}, much as the phonon do in the conventional pairing mechanism. As high-temperature superconductivity, solid or non-solid materials are referred to, the superconductivity, unlike conventional superconductor, does not come from the electron-phonon interaction. Most of the time it occurs in not metallic, but ceramic materials\cite{7,11}. Although it seems responsible for the pair formation for the superconductivity, but one suggests unconventional electron mating mechanisms. Superconductors with high superconducting transition ($T_{\text{c}}$) have unusual properties that cannot be explained by conventional Fermi liquid or BCS theory of superconductivity. One major open question is the pairing mechanism in unconventional superconductors. Other works, due to Yang and Kong\cite{8,13} have shown that, by the mechanism of charge fluctuations for the cuprate oxide superconductors all the current carriers take part in the superconductivity and exhibit the proper superconducting window of high $T_{\text{c}}$ superconductors reasonably. Since the mechanism of charge fluctuations is associated with the localized property of electrons, the two-band model has been shown to be satisfactory. Yu et al. generalized this idea to the superconductivity for the non-cuprate compound $BaPb_{\{1-x\}} BixO_{3}$ \cite{14,15}. Basing on the singly occupied picture and the charge fluctuations, the effective Hamiltonian of two-band Hubbard model is derived by the four-order degeneracy perturbation theory. In terms of the obtained Green function equation of motion, one can get a curve of $T_{\text{c}}-\delta$ which is consistent with experiment data qualitatively and quantitatively, with $T_{\text{c}} \approx 30 K$. The results show that, for the Cooper pairing, the charge fluctuation mechanism is more important than the phonon mechanism.

Motivated by Ashocroft’s 2004 suggestion \cite{16,17,18} that high temperature superconductivity should exist in hydrogen-rich compounds under high pressure, an intensive experimental and theoretical search got underway \cite{19}. Unconventional superconductors react to pressure differently than conventional one. There are significant difference between the Cooper pairs of different superconductor type. The key difference between other unidentified superconducting objects and hydrides is that there is a strong expectation that hydrides under pressure are high temperature, because it is predicted by the conventional theory. However, for the hydrides under high pressures, where only a few select groups have ability to do these experiments, and the samples are extremely small and difficult to handle, it is much harder to confirm or rule out such indication. In addition, experiments in hydrides under high pressure suffer from a lack of reproducibility. There is no guarantee that when an experiment is repeated, even by the same group, the same compound is synthesized \cite{20,21,22,23,24}.

The current view, motivated by the experimental findings in recent years that there are many superconducting materials that do not fit the BCS framework, holds that there is no such thing as the theory of superconductivity \cite{9,19,25,26,27,28,29,30,31}. There is a theory of superconductivity to describe the superconducting elements and simple compound (the conventional theory), another theory to describe the high $T_{\text{c}}$ cuprates, yet another one for the pnictides, another for heavy fermion materials, another for $\mathrm{Sr}_{2} \mathrm{RuO}_{3}$, another for $\mathrm{SrTiO}_{3}$, another for bismuthates, another for hydrides, etc. If any, the right theories among many proposed candidates are to describe each of these numerous other classes of superconductors, and in particular, how many different theories are needed to describe the superconducting materials \cite{10,32,33,34,35,36,37,38}. In fact, there may be some classes of superconductors where particular characteristic of the materials are closely intertwined with their superconductivity giving rise to some different physics, yet the underlying mechanism for superconductivity would not be different than for any other material. If such a unifying theory exists, with a pairing mechanism that under favorable conditions can give rise to the $140 \mathrm{~K}$ superconductivity of mercury-barium-calcium copper oxide, it would be very surprising that under less favorable conditions, the same pairing mechanism could also give rise to the $7 \mathrm{~K}$ superconductivity of $\mathrm{Pb}$ \cite{24}. From this reason, that a single mechanism of superconductivity is by far the more likely scenario to describe superconductivity in nature. The original BCS Hamiltonian $\bar{H}=\sum_{k} \epsilon_{k}\left(C_{k \uparrow}^{\dagger} C_{-k \downarrow}^{\dagger}+C_{-k \downarrow}^{\dagger} C_{-k \downarrow}\right)+H_{int}$ and their pairing idea are faithful and correct, in which the attractive two-mode interaction $H_{\text{int}}=-V \sum_{k, k^{\prime}} C_{k^{\prime}\uparrow}^{\dagger} C_{-k^{\prime} \downarrow}^{\dagger} C_{-k \downarrow} C_{k \uparrow}$ is crucial in the formation for superconducting state. For weak coupling, BCS theory is a successful one, but for the high-temperature superconductors, their theory is a failure: 1) Since the electron-electron interaction mediated by the electron-phonon interaction $V_{\text{el-el}} \ll \hbar \omega_{\text{D}}(\approx 0.01 ~\text{eV} )$, the phonon Cooper pairing is not stable in the high-temperature superconductors; 2) The high-temperature superconductivity is closely related to the strong coupling, the energy integral upper limit $\hbar \omega_{0} \gg \hbar \omega_{\text{D}}$, the cut-off frequency $\hbar \omega_{0}$ is a open difficulty to solve up to now; 3) For the BCS theory, starting from $-V\sum_{k,k^{\prime}}C_{k^{\prime}\uparrow}^{\dagger}C_{-k^{\prime}\downarrow}^{\dagger}C_{-k\downarrow}C_{k\uparrow}$ , using Wick contraction and $\sum_{k}\langle C_{-k\downarrow}C_{k\uparrow} \rangle=\sum_{k^{\prime}}\langle C_{k^{\prime}\uparrow}^{\dagger}C_{-k^{\prime}\downarrow}^{\dagger} \rangle$, in this way, a red carpet is used to conceal the non-physical terms $ C_{-k\downarrow}C_{k\uparrow} \langle C_{-k^{\prime}\downarrow}C_{k^{\prime}\uparrow} \rangle$ and $ C_{k\uparrow}^{\dagger}C_{-k\downarrow}^{\dagger} \langle C_{k^{\prime}\uparrow}^{\dagger}C_{-k^{\prime}\downarrow}^{\dagger} \rangle$, and the Cooper pair scattering processes $C_{-k \downarrow}C_{k \uparrow}\rightarrow C_{k^{\prime}\uparrow}^{\dagger}C_{-k^{\prime}\downarrow}^{\dagger}$ with $k^{\prime} \neq k$. As this result, the energy gap $\Delta$ and the cut-off frequency $\hbar \omega_0$ cannot be given as unique solution. 4) The strong coupling theory had some unsolved problem from the very time of its creation in physics; 5) The superconductivity is a phase coherent quantum phenomenon, as a most critical correction the coherent interaction and action-counteraction are not considered.

As we shown, for the Watson’s two helix model of DNA, the base pairing is originated from the stacking force $\widetilde{V}_{\text{st}}$ but not the hydrogen bond. Namely, for the superconductors, the pairing of $C_{k \uparrow}$ and $C_{-k \downarrow}$ is also originated from the local fluctuation stacking force but not from the electron-electron interaction $V_{\text{el-el}}$(e-p) mediated by the electron-phonon interaction.

This paper, in Sec.2, as the model Hamiltonian we give the microscopic principle of strong local fluctuation stacking force, which is related to the pairing mechanism for two-mode electrons $C_{k \uparrow}$ and $C_{-k \downarrow}$; by use of quaternions method, the complex Bogoliubov transformation is obtained. Subsequently, in Sec.3 we consider the one-electron coherent state and the two-mode electron coherent state,and set forth the strong local variational theory. Finally, in Sec.4, based on the principles of coherent interaction and action-counteraction\cite{39}, we have constructed the energy variational extremum equation and the energy gap equation, and give their self-consistent solutions further. The paper is concluded in Sec.5 with discussion.
\section{Superconducting model Hamiltonian with local fluctuation stacking force pairing , microscopic principle of strong local quantum fluctuation mechanism for local staking force, and the complex Bogoliubov transformation}
\subsection{Superconducting model Hamiltonian,microscopic principle of strong local quantum fluctuations for local stacking force}

Nature obeys the principle of pairing, which is an untenable principle of natural philosophy. Therefore, we should consider the two-mode scattering interaction
\begin{equation}
	H_{\text{int}}=-V\sum_{k,k^{\prime}}C_{k^{\prime}\uparrow}^{\dagger}C_{-k^{\prime}\downarrow}^{\dagger}C_{-k\downarrow}C_{k\uparrow}.
\end{equation}

Looking at $V\langle\Phi|C_{k^{\prime} \uparrow}^{\dagger} C_{-k^{\prime} \downarrow}^{\dagger} C_{-k \downarrow} C_{k \uparrow}| \Phi\rangle$, since $C_{k \uparrow} C_{-k \downarrow} \rightarrow C_{k^{\prime} \uparrow}^{\dagger} C_{-k^{\prime} \downarrow}^{\dagger}$, it does not to treat in the Hartree-Fock approximation and is a quite new physics. It is, of course, of crucial importance for the whole of superconductivity theory.

For the BCS superconductivity, the paring of $C_{k \uparrow}$ and $C_{-k \downarrow}$ is attributed to the electron-electron interaction $V_{\text{el-el}}$(e-p)  mediated by the lattice phonons, the phonon Cooper pair. However, the phonon energy $\hbar \omega_{\text{D}} \approx 0.01~\text{eV} $, but $V_{\text{el-el}} \text{(e-p)} \ll \hbar \omega_{\text{D}}$, as this result, the phonon Cooper pairing is not stability in the high-$T_{\text{c}}$ superconductors. Consulting the pairing mechanism of the Watson’s two helix model of DNA, we first set forward the strong local stacking force pairing mechanism of two-mode electrons $C_{k}$ and $C_{-k}$ caused by the strong local fluctuation action, with the tying pair $C_{-k\downarrow}C_{k\uparrow}e^{ik\cdot r}$ and two straight pairing orbits.

When electrons $C_{k \uparrow}$ and $C_{-k \downarrow}$ move in opposite directions with \textit{v} and -\textit{v} in two parallel straight pairing orbits, the kinetic energy $E_{k}=\frac{1}{2} \frac{m\times m}{m+m}(2v)^{2}=\frac{p^2}{m}=\frac{p^2}{2m}+\frac{p^2}{2m}$, the kinetic energy of two unpaired electrons $C_{k \uparrow}$ and $C_{-k \downarrow}$. However, when electrons $C_{k \uparrow}$ and $C_{-k \downarrow}$ have combined to form a Cooper pairing, the composite elementary particle, at that time two electrons $C_{k \uparrow}$  and $C_{-k \downarrow}$ are in the local stacking state. Considering the local fluctuation of $x=|x(C_{k \uparrow})-x(C_{-k \downarrow})|$ and $\langle \Phi(x)|\Delta^{2}x|\Phi(x)\rangle=\langle \Phi(x)|x^{2}|\Phi(x)\rangle-\langle \Phi(x)|x|\Phi(x)\rangle^{2}$ with $\Phi(x)=\left(\frac{\lambda}{\sqrt{\pi}}\right)^\frac{1}{2}e^{-\frac{1}{2} \lambda^2 x^2}$, we can obtain the local fluctuation of $x$ as
\begin{equation}
\label{Eq.2}
        \langle \Delta^{2}x \rangle=\frac{1}{4 \lambda^2}-\frac{1}{4 \pi \lambda^2}=\frac{1}{4} \left( 1-\frac{1}{\pi} \right)\frac{1}{\lambda^2}.
\end{equation}

Obviously, when $\lambda \rightarrow \infty$, $\Delta^2 x \rightarrow 0$, two electrons $C_{k \uparrow}$ and $C_{-k \downarrow}$ will occur in a strongly local stacking state. On the basis of local stacking mechanism, since $s_1 \cdot s_2=-\frac{3}{4}$, we will see the attractive interaction between two electrons $C_{k \uparrow}$ and $C_{-k \downarrow}$ caused by the local stacking action as follows
\begin{equation}
\label{Eq.3}
    V_{\text{st}}=\langle \Phi(r)|\frac{p^2}{2m}+\frac{p^2}{2m}|\Phi(r)\rangle s_{1}(\uparrow)\cdot s_{2}(\downarrow)=-\frac{9}{4}\lambda^{2} r_{0}^{2},
\end{equation}
where $\Phi(r)=\left(\frac{\lambda}{\sqrt{\pi}}\right)^\frac{3}{2}e^{-\frac{1}{2} \lambda^2 r^2}$, $r_{0}^2=\frac{\hbar^2}{2m}$. Clearly, when $\lambda=0$, two electrons $C_{k \uparrow}$ and $C_{-k \downarrow}$ are in a itinerant state. Obviously, the attractive interaction $V_{\text{st}}$ is originated from the local stacking action without the local fluctuations.

Because of the scattering interaction $-V \sum_{k,k^{\prime}} C_{k^{\prime} \uparrow}^{\dagger} C_{-k^{\prime} \downarrow}^{\dagger} C_{-k \downarrow} C_{k \uparrow}$, it transfers the two-mode state from $\left|k \uparrow,-k\downarrow \right\rangle$ to $\left|k^{\prime} \uparrow,-k^{\prime}\downarrow \right\rangle$, this process causes the local quantum fluctuations to occur. At the same time, the nonlinear action of the two-mode squeezed coherent state $|Z_{\uparrow \downarrow}\rangle= e^{\sum_{k}Z_{\uparrow \downarrow} \left(C_{k \uparrow}^{\dagger} C_{-k\downarrow}^{\dagger}-C_{-k\downarrow} C_{k \uparrow}\right)}|\Phi^{(2)}(0)\rangle$ further increases the local quantum fluctuations. Therefore based on the interaction of electrons $C_{k \uparrow}$ and $C_{-k \downarrow}$ with the local quantum fluctuations, noting that
\begin{equation}
\begin{split}
\langle\Phi(r)|\Delta^2 p|\Phi(r)\rangle=&\langle \Phi(r) |p^2| \Phi(r) \rangle-\langle \Phi(r) |p| \Phi(r) \rangle^2\\
=&\hbar^2\left(\frac{3}{2}\lambda^2+\frac{4}{\pi}\lambda^2\right),
\end{split}
\label{Eq.4}
\end{equation}
we obtain for the attractive interaction between electrons $C_{k\uparrow}$ and $C_{-k \downarrow}$ caused by the local quantum fluctuation principle as
\begin{equation}
\label{Eq.5}
    \tilde{V}_{\text{st}}=3\lambda^2 r_0^2 \left(1+\frac{8}{3\pi}\right)s_{1}(\uparrow)\cdot s_{2}(\downarrow)=-\frac{9}{4}\left(1+\frac{8}{3\pi}\right)\lambda^2 r_{0}^2.
\end{equation}

Let $\tilde{E}_k=\frac{p_1^2}{2m}+\frac{p_2^2}{2m}=\hbar^2 k_1^2+\hbar^2 (k_0-k_1)^2$, from $\frac{d\tilde{E}_k}{dk_1}=0$ we have $k_1=k_0-k_1=k_2$. That is, when $s_1(\uparrow)\cdot s_2(\downarrow)=-\frac{3}{4}$ and $\vec{k}_1=-\vec{k}_2=\vec{k}$, both $\tilde{E}_k$ and $\tilde{V}_{\text{st}}$ have maximum values.

In virtue of the local fluctuation principle, $|\tilde{V}_{\text{st}}|>|V_{\text{st}}|$. From Eq.~(\ref{Eq.2}) and Eq.~(\ref{Eq.5}) we see that, for $\lambda \rightarrow 0$, $\langle \Delta^2 x \rangle \rightarrow \infty$, two electrons $C_{k\uparrow}$ and $C_{-k \downarrow}$ are very far from each other and $\tilde{V}_{\text{st}} \rightarrow 0$, that is, the itinerant electron state; but for $\lambda \rightarrow \infty$, $\langle \Delta^2 x \rangle \rightarrow 0$, $C_{k \uparrow}$ and $C_{-k \downarrow}$ are very close to each other and $\tilde{V}_{\text{st}} \rightarrow \infty$, that is, the strongly local stacking state.

From Eq.~(\ref{Eq.3}) and Eq.~(\ref{Eq.5}), we can see that, since $p^{2}=\hbar^{2} k^{2}$, for a given coupling strength, so then $\frac{k^{2}}{\lambda^{2}}$ is a constant. Obviously $k^{2}$ increases with the attractive local stacking force, at the same time $\lambda^2$ also will increase to a larger value since $\frac{k^{2}}{\lambda^{2}}$ is a constant. That is shown, with time evolution, $k^{2}$ becomes larger, $\lambda^{2}$ becomes larger, at the end the strong local fluctuation stacking force will develop to a very large value, $\tilde{V}_{\text{st}} \gg \hbar \omega_{\text{D}} \gg V_{\text{el-el}}(\text{e-p})$.

\subsection{Quaternions method and complex Bogoliubov transformation for the fermion systems}
A quaternion $Q$ is a linear combination $a I+b i_{1}+c i_{2}+d i_{3}$ where $a, b, c, d$ are real numbers with $I=(1,0,0,0), i_{1}=(0,1,0,0), i_{2}=$ $(0,0,1,0), i_{3}=(0,0,0,1)$.

The sum of quaternions is the usual componentwise sum and multiplication is defined so that $(1,0,0,0)$ is the identity and for fermion systems, $i_{1}, i_{2}, i_{3}$ satisfy
\begin{equation}
\label{Eq.6}
i_{1}^{2}=i_{2}^{2}=i_{3}^{2}=i_{1} i_{2} i_{3}=-1.
\end{equation}
It follows from Eq.~(\ref{Eq.6}) that
\begin{equation}
\label{Eq.7}
\begin{array}{c}
i_{1}i_{2}=-i_{2}i_{1}=i_{3},\\
i_{2}i_{3}=-i_{3}i_{2}=i_{1},\\
i_{3}i_{1}=-i_{1}i_{3}=i_{2}.
\end{array}
\end{equation}

To realize the quaternion multiplication for fermion systems and map to $SU(2)$, we introduce
\begin{equation}
\label{Eq.8}
\sigma_{1}=\left[\begin{array}{ll}
0 & 1 \\
1 & 0
\end{array}\right], \sigma_{2}=\left[\begin{array}{cc}
0 & -i \\
i & 0
\end{array}\right], \sigma_{3}=\left[\begin{array}{cc}
1 & 0 \\
0 & -1
\end{array}\right].
\end{equation}

Noting that $i \sigma_{1} i \sigma_{3} i \sigma_{2}=i \sigma_{3} i \sigma_{2} i \sigma_{1}=i \sigma_{2} i \sigma_{1} i \sigma_{3}=-\left[\begin{array}{ll}1 & 0 \\ 0 & 1\end{array}\right]$, then the quaternions $(a, b, c, d)$ map to $S U(2)$ as following
\begin{equation}
\label{Eq.9}
\begin{array}{c}
Q_{F}^{(1)}=a I+i \sigma_{1} b+i \sigma_{3} c+i \sigma_{2} d,  \\
Q_{F}^{(2)}=a I+i \sigma_{3} b+i \sigma_{2} c+i \sigma_{1} d,  \\
Q_{F}^{(3)}=a I+i \sigma_{2} b+i \sigma_{1} c+i \sigma_{3} d.
\end{array}
\end{equation}

Later we will use the map
\begin{equation}
\label{Eq.10}
\begin{aligned}
(a, b, c, d) \rightarrow Q_{F}&=a I+i \sigma_{3} b-i \sigma_{2} c-i \sigma_{1} d \\
&=\left[\begin{array}{cc}
a+i b & -(c+i d) \\
c-i d & a-i b
\end{array}\right],
\end{aligned}
\end{equation}
from this we find the complex Bogoliubov transformations as
\begin{equation}
\label{Eq.11}
\left[\begin{array}{c}
\alpha_{k} \\
\alpha_{-k}^{\dagger}
\end{array}\right]=\left[\begin{array}{cc}
a+i b & -(c+i d) \\
c-i d & a-i b
\end{array}\right]\left[\begin{array}{c}
C_{k} \\
C_{-k}^{\dagger}
\end{array}\right],
\end{equation}
and its reverse transformation is
\begin{equation}
\label{Eq.12}
\left[\begin{array}{c}
C_{k} \\
C_{-k}^{\dagger}
\end{array}\right]=\left[\begin{array}{cc}
a-i b & c+i d \\
-(c-i d) & a+i b
\end{array}\right]\left[\begin{array}{c}
\alpha_{k} \\
\alpha_{-k}^{\dagger}
\end{array}\right],
\end{equation}
with $\left(a^{2}+b^{2}\right)+\left(c^{2}+d^{2}\right)=1$.

Based on above crucial attractive interaction between the two-mode electrons $C_{k \uparrow}$ and $C_{-k \downarrow}$, the superconducting model Hamiltonian with $V=V_{\text{el-el}}(\text{e-p})+\tilde{V}_{\text{st}}$ is 
\begin{equation}
\label{Eq.13}
\begin{aligned}
 H&=\sum_{k} \epsilon_{k}\left(C_{k \uparrow}^{\dagger} C_{k \uparrow}+C_{-k \downarrow}^{\dagger} C_{-k \downarrow}\right)\\
 &-V \sum_{k, k^{\prime}}\left(C_{k^{\prime} \uparrow}^{\dagger} C_{-k^{\prime} \downarrow}^{\dagger} C_{-k \downarrow} C_{k \uparrow}\right).  
 \end{aligned}
\end{equation}

(Later, $C_{k \uparrow}=C_{k}, C_{-k \downarrow}=C_{-k}$.) After the Wick contraction once, by the first order approximation, we have
\begin{equation}
\label{Eq.14}
\begin{aligned}
H &\approx \sum_{k} \epsilon_{k}\left(C_{k}^{\dagger} C_{k}+C_{-k}^{\dagger} C_{-k}\right)  \\
& -V \sum_{k, k^{\prime}}\left[\langle C_{-k} C_{k}\rangle C_{k^{\prime}}^{\dagger} C_{-k^{\prime}}^{\dagger}+\langle C_{k^{\prime}}^{\dagger} C_{-k^{\prime}}^{\dagger}\rangle C_{-k} C_{k}\right] \\
& +V \sum_{k, k^{\prime}}\langle C_{k^{\prime}}^{\dagger} C_{-k^{\prime}}^{\dagger}\rangle\langle C_{-k} C_{k}\rangle .
\end{aligned}\end{equation}

Introducing of the energy gap parameter
\begin{equation}
\label{Eq.15}
\Delta=V \sum_{k^{\prime}}\langle C_{-k^{\prime}} C_{k^{\prime}}\rangle, \Delta^{*}=V \sum_{k^{\prime}}\langle C_{k^{\prime}}^{\dagger} C_{-k^{\prime}}^{\dagger}\rangle,
\end{equation}
the Hamiltonian~(\ref{Eq.14}) takes the approximate form (BCS Hamiltonian) as follow
\begin{equation}
\label{Eq.16}
\begin{aligned}
H&=\sum_{k} \epsilon_{k}\left(C_{k}^{\dagger} C_{k}+C_{-k}^{\dagger} C_{-k}\right)\\
&-\Delta \sum_{k}\left(C_{k}^{\dagger} C_{-k}^{\dagger}+C_{-k} C_{k}\right)+\frac{\Delta^{2}}{V} .
\end{aligned}\end{equation}

To avoid exchanging the Cooper pairing wih other Cooper pairs, the tying Cooper pair will replace the itinerant Cooper pair for increasing the ground state energy gap greatly, we rewrite the Hamiltonian as
\begin{equation}
\label{Eq.17}
\begin{aligned}
H&=\sum_{k} \epsilon_{k}\left(C_{k}^{\dagger} C_{k}+C_{-k}^{\dagger} C_{-k}\right)\\
&-\Delta \sum_{k}\left(C_{k}^{\dagger} C_{-k}^{\dagger} e^{-i k \cdot r}+C_{-k} C_{k} e^{i k \cdot r}\right)+
\frac{\Delta^{2}}{V}.
\end{aligned}
\end{equation}

Considering $[C_{k}, H]=\epsilon_{k} C_{k}-\Delta C_{-k}^{\dagger} e^{-i k \cdot r}$, $[C_{-k}^{\dagger}, H]=-\epsilon_{k} C_{-k}^{\dagger}-\Delta C_{k} e^{i k \cdot r}$, we obtain
\begin{equation}
\label{Eq.18}
\begin{aligned}
\left[\alpha_{k}, H\right]&=(a+i b)\left(\epsilon_{k} C_{k}-\Delta C_{-k}^{\dagger} e^{-i k \cdot r}\right)\\
&+(c+i d)\left(\epsilon_{k} \epsilon_{-k}^{\dagger}+\Delta C_{k} e^{i k \cdot r}\right) .
\end{aligned}
\end{equation}

On the other hand, for the $\alpha_{k}$-representation, $H=\sum_{k} E_{k}\left(\alpha_{k}^{\dagger} \alpha_{k}+\right.$ $\left.\alpha_{-k}^{\dagger} \alpha_{-k}\right)$, thus we get
\begin{equation}
\label{Eq.19}
\left[\alpha_{H}, H\right]=E_{k}\left[(a+i b) C_{k}-(c+i d) C_{-k}^{\dagger}\right].
\end{equation}

By combining Eq.~(\ref{Eq.18}) with Eq.~(\ref{Eq.19}), we can arrive at
\begin{equation}
\label{Eq.20}
\begin{array}{c}
\left(\epsilon_{k}-E_{k}\right)(a+ib)+\Delta(c+id)e^{ik\cdot r}=0,\\
-\Delta e^{-ik\cdot r}(a+ib)+\left(\epsilon_{k}+E_{k}\right)(c+id)=0,
\end{array}
\end{equation}
and from Eq.~(\ref{Eq.20}), one finds $E_{k}^{2}=\epsilon_{k}^{2}+\Delta^{2}$. Furthermore, combining $\left(\epsilon_{k}-E_{k}\right)(a+i b)+\Delta(c+i d) e^{i k \cdot r}=0$ with $\left(\epsilon_{k}-\right.$ $\left.E_{k}\right)(a-i b)+\Delta(c-i d) e^{-i k \cdot r}=0$ and $\left(a^{2}+b^{2}\right)+\left(c^{2}+d^{2}\right)=1$, we yield
\begin{equation}
\label{Eq.21}
a^{2}+b^{2}=\frac{E_{k}+\epsilon_{k}}{2 E_{k}}, c^{2}+d^{2}=\frac{E_{k}-\epsilon_{k}}{2 E_{k}},\end{equation}
and
\begin{equation}
\label{Eq.22}
a=\frac{\Delta}{2 E_{k}}, b=\frac{E_{k}+\epsilon_{k}}{2 E_{k}}, c=\frac{E_{k}-\epsilon_{k}}{2 E_{k}}, d=\frac{\Delta}{2 E_{k}} .\end{equation}

\section{One-electron coherent state and two-mode electron coherent state, strong local variational approach}
In accordance with the two-mode attractive action, as this result it will simultaneously induce the two-mode coherent action $H_{\text{c}}^{(2)}=$ $\sum_{k} i \hbar\left(\zeta C_{k \uparrow}^{\dagger} C_{-k \downarrow}^{\dagger}-\zeta^{*} C_{-k \downarrow} C_{k \uparrow}\right)$. In Schrödinger picture, state $\left|\Phi^{(2)}(t)\right\rangle$ evolves in time according to the rule
\begin{equation}
\label{Eq.23}
\begin{aligned}
\left|\Phi^{(2)}(t)\right\rangle&=e^{-\frac{i}{\hbar} H_{\text{c}}^{(2)} t}\left|\Phi^{(2)}(0)\right\rangle \\
&=e^{\sum_{k}\left(\zeta C_{k \uparrow}^{\dagger} C_{-k \downarrow}^{\dagger}-\zeta^{*} C_{-k \downarrow} C_{k \uparrow}\right) t}\left|\Phi^{(2)}(0)\right\rangle.
\end{aligned}\end{equation}

Hence the state of the field generated by coherent action $H_{\text{c}}^{(2)}$ is the two-mode electron coherent state, the Cooper pair squeezed state,
\begin{equation}
\label{Eq.24}
\begin{aligned}
\left|Z_{\uparrow \downarrow}\right\rangle&=S\left(Z_{\uparrow \downarrow}\right)\left|\Phi^{(2)}(0)\right\rangle,\\ S\left(Z_{\uparrow \downarrow}\right)
&=e^{\sum_{k} Z_{\uparrow \downarrow}\left(C_{k \uparrow}^{\dagger} C_{-k \downarrow}^{\dagger}-C_{-k \downarrow} C_{k \uparrow}\right)},
\end{aligned}
\end{equation}
where $Z_{\uparrow \downarrow}$ is the two-mode squeezing parameter. In line with two-mode squeezed transformation, we can obtain
\begin{equation}
\label{Eq.25}
\begin{aligned}
S C_{k} S^{\dagger}&=\tilde{C}_{k}=C_{k} \cos Z-C_{-k}^{\dagger} \sin Z, \\
S C_{-k}^{\dagger} S^{\dagger}&=\tilde{C}_{-k}^{\dagger}=C_{-k}^{\dagger} \cos Z+ C_k \sin Z,
\end{aligned}
\end{equation}
and the reverse transformation
\begin{equation}
\label{Eq.26}
\begin{aligned}
C_{k}&=\tilde{C}_{k} \cos Z+\tilde{C}_{-k}^{\dagger} \sin Z, \\
C_{-k}^{\dagger}&=\tilde{C}_{-k}^{\dagger} \cos Z-\tilde{C}_{k} \sin Z.
\end{aligned}
\end{equation}
On the other hand, because of the phase coherence between the electrons among the system, it will induce the one-electron coherent action at the same time, $H_{\text{c}}^{(1)}=\sum_{k} i \hbar\left(f_{1} C_{k}^{\dagger}-f_{1}^{*} C_{k}\right)$. Consequently, state $|\Phi^{(1)}(t)\rangle=e^{-\frac{i}{h} H_{\text{c}}^{(1)} t}|\Phi^{(1)}(0)\rangle$ of the field generated by $H_{\text{c}}^{(1)}$ will finally evolve into the one-electron coherent state
\begin{equation}
\label{Eq.27}
|\eta\rangle=D(\eta)|0\rangle_{\text{c}}, D(\eta)=e^{\sum_{k}\left(\eta C_{k}^{\dagger}-\eta^{*} C_{k}\right)}.\end{equation}

The one-fermion coherent state is the eigen state of fermion operator
\begin{equation}
\label{Eq.28}
C_{k \uparrow}\left|\eta_{+}\right\rangle=\eta_{+}\left|\eta_{+}\right\rangle, C_{-k \downarrow}\left|\eta_{-}\right\rangle=\eta_{-}\left|\eta_{-}\right\rangle \text {. }\end{equation}

Because $\left(C_{k}^{2}\right)=\left(C_{k}^{\dagger}\right)^{2}=0$, the eigen-values $\eta$ and $\eta^{*}$ satisfy the Grassman algebra,
\begin{equation}
\label{Eq.29}
\eta^{2}=0, \eta^{* 2}=0, \eta \eta^{*}+\eta^{*} \eta=0,\end{equation}
with this reason, we have
\begin{equation}
\label{Eq.30}
\begin{aligned}
\left|\eta_{+}\right\rangle&=e^{C_{k \uparrow}^{\dagger} \eta_{+}}|0\rangle=D\left(\eta_{+}\right)|0\rangle, \\~\left|\eta_{-}\right\rangle&=e^{C_{-k \downarrow}^{\dagger} \eta_{-}}|0\rangle=D\left(\eta_{-}\right)|0\rangle .
\end{aligned}
\end{equation}

By means of $D\left(\eta_{+}\right)$and $D\left(\eta_{-}\right)$, noting the Grassman algebra, we find the displaced transformations for $\left(C_{k}, C_{-k}\right)$ and $\left(C_{k}^{\dagger}, C_{-k}^{\dagger}\right)$ as
\begin{equation}
\label{Eq.31}
\begin{array}{c}
e^{C_{k}^{\dagger}\eta_{+}}C_{k}e^{-C_{k}^{\dagger}\eta_{+}}=C_{k}-2S_{+}\eta_{+},\\
e^{\eta_{+}^{*}C_{k}}C_{k}^{\dagger}e^{-\eta_{+}^{*}C_{k}}=C_{k}^{\dagger}-2S_{+}\eta_{+}^{*},\\
e^{C_{-k}^{\dagger}\eta_{-}}C_{-k}e^{-C_{-k}^{\dagger}\eta_{-}}=C_{-k}-2S_{-}\eta_{-},\\
e^{\eta_{-}^{*}C_{-k}}C_{-k}^{\dagger}e^{-\eta_{-}^{*}C_{-k}}=C_{-k}^{\dagger}-2S_{-}\eta_{-}^{*},
\end{array}
\end{equation}
where
\begin{equation}
\label{Eq.32}
2 S_{+}=C_{k} C_{k}^{\dagger}-C_{k}^{\dagger} C_{k}, 2 S_{-}=C_{-k} C_{-k}^{\dagger}-C_{-k}^{\dagger} C_{-k}. 
\end{equation}

BCS Hamiltonian (\ref{Eq.14}) is based on the $-V\sum_{k,k^{\prime}}C_{k^{\prime}\uparrow}^{\dagger}C_{-k^{\prime}\downarrow}^{\dagger}C_{-k\downarrow}C_{k\uparrow}$ , using Wick contraction and $\sum_{k}\langle C_{-k\downarrow}C_{k\uparrow} \rangle=\sum_{k^{\prime}}\langle C_{k^{\prime}\uparrow}^{\dagger}C_{-k^{\prime}\downarrow}^{\dagger} \rangle$, in this way, a red carpet is used to conceal the $\langle C_{-k\downarrow}C_{k\uparrow}\rangle \neq \langle C_{k^{\prime}\uparrow}^{\dagger}C_{-k^{\prime}\downarrow}^{\dagger} \rangle$ and the Cooper pair scattering processes $C_{-k \downarrow}C_{k \uparrow}\rightarrow C_{k^{\prime}\uparrow}^{\dagger}C_{-k^{\prime}\downarrow}^{\dagger}$ with $k^{\prime} \neq k$. As this result, the energy gap $\Delta$ and the cut-off frequency $\hbar \omega_0$ cannot be given as unique solution. Noting $ V  \sum\limits_k {{C_{ - k}}} {C_k}\sum\limits_{k'} { \langle C_{k'}^ {\dagger} } C_{ - k'}^ {\dagger}  \rangle $ or $ V  \sum\limits_{k'} {C_{k'}^ {\dagger} } C_{ - k'}^ {\dagger} \sum\limits_k { \langle {C_{ - k}}} {C_k} \rangle $ in Eq.~(\ref{Eq.14}), there will be two different evolution cases. For $ V  \sum\limits_k {{C_{ - k}}} {C_k}\sum\limits_{k'} {\langle C_{k'}^ {\dagger} } C_{ - k'}^ {\dagger}  \rangle$, we have two evolution cases:  
\begin{enumerate}
	\item $ V  \sum\limits_k {{C_{ - k}}} {C_k}\sum\limits_{k'} { \langle C_{k'}^ {\dagger} } C_{ - k'}^ {\dagger}  \rangle = \Delta \sum\limits_k {{C_{ - k}}} {C_k}$.
	\item $ V  \sum\limits_k {{C_{ - k}}} {C_k}\sum\limits_{k'} { \langle C_{k'}^ {\dagger} } C_{ - k'}^ {\dagger}  \rangle  =  V  \sum\limits_k {{C_{ - k}}} {C_k}\left[ { \cdots  +  \langle C_{ - k'}^ {\dagger} C_{ - k'}^ {\dagger}  \rangle  \cdots  +  \langle C_k^ {\dagger} C_{ - k}^ {\dagger}  \rangle  \cdots } \right] = V  \sum\limits_k {{C_{ - k}}} {C_k} \langle C_k^ {\dagger} C_{ - k}^ {\dagger}  \rangle  =  V  \sum\limits_k { \langle{C_{ - k}}} {C_k} \rangle {C_{ - k}}{C_k} $.
\end{enumerate}

The reasons are as follows. 1) The physical process of $C_{-k}C_k$ and its statistical average $\langle C_{k}^{\dagger}C_{-k}^{\dagger}\rangle=\langle C_{-k}C_k\rangle$ have definite phase relationship and cause phase coherence; 2) Two identical physical events $C_{-k}C_k$ and $\langle C_{-k}C_k\rangle $ are compatible; Therefore,  $C_{-k}C_k \langle C_{k}^{\dagger}C_{-k}^{\dagger}\rangle$ is an objective existence. However, $C_{-k}C_k$ and $\langle C_{k'}^{\dagger}C_{-k'}^{\dagger}\rangle $ have no phase coherence, they are not the compatable events, so then, $C_{-k}C_k\langle C_{k'}^{\dagger}C_{-k'}^{\dagger}\rangle$ is not objective.

Similarly, for $ V  \sum\limits_{k'} {C_{k'}^ {\dagger} } C_{ - k'}^ {\dagger} \sum\limits_k {\left\langle {{C_{ - k}}{C_k}} \right\rangle } $, we also have two evolution cases:

\begin{enumerate}
	\item $ V  \sum\limits_{k'} {C_{k'}^ {\dagger} } C_{ - k'}^ {\dagger} \sum\limits_k {\left\langle {{C_{ - k}}{C_k}} \right\rangle }  = \sum\limits_{k'} {C_{k'}^ {\dagger} } C_{ - k'}^ {\dagger} \Delta  = \Delta \sum\limits_k {C_k^ {\dagger} C_{ - k}^ {\dagger} } $.
	\item $ V  \sum\limits_{k'} {C_{k'}^ {\dagger} } C_{ - k'}^ {\dagger} \sum\limits_k {\left\langle {{C_{ - k}}{C_k}} \right\rangle }  =  V  \sum\limits_{k'} {C_{k'}^ {\dagger} } C_{ - k'}^ {\dagger} \left\langle {{C_{ - k'}}{C_{k'}}} \right\rangle  =  V  \sum\limits_k {\left\langle {{C_{ - k}}{C_k}} \right\rangle C_k^ {\dagger} C_{ - k}^ {\dagger} } $.
\end{enumerate}

With above results, we obtain another Hamiltonian of superconductors describing the superconducting transition
\begin{equation}
\label{Eq.33}
\begin{aligned}
H_{S}&=\sum_{k} \epsilon_{k}\left(C_{k}^{\dagger} C_{k}+C_{-k}^{\dagger} C_{-k}\right)
-V \sum_{k}\langle C_{-k} C_{k}\rangle\\
&\left(C_{k}^{\dagger} C_{-k}^{\dagger} e^{-i k \cdot r}+C_{-k} C_{k} e^{i k \cdot r}\right)
+\frac{\Delta^{2}}{V}
\end{aligned}\end{equation}
with $\langle C_{-k} C_{k}\rangle \approx \frac{\Delta}{2 E_{k}}$. In this manner, by combining $H_{\text{BCS}}$ and $H_S$ with each other, the unknowns $\Delta$ and $\hbar \omega_0$ can be given as unique solution.

As it were, for the second quantization representation
\begin{equation}
\label{Eq.34}
\begin{aligned}
\int d x \psi^{\dagger}(x) \hat{p} \psi(x)&=\sum_{k} \hbar k C_{k}^{\dagger} C_{k}, \\
\int d x \psi^{\dagger}(x) \frac{\hat{p}^{2}}{2 m} \psi(x)&=\sum_{k} \frac{\hbar^{2} k^{2}}{2 m} C_{k}^{\dagger} C_{k}\\
&=\sum_{k} \epsilon_{k} C_{k}^{\dagger} C_{k},
\end{aligned}\end{equation}
where $\psi(x)=\sum_{k} C_{k} \phi(x)$, with $\phi(x)=\frac{1}{\sqrt{\Omega}} e^{i k \cdot x} \chi_{\frac{1}{2}}(z)$.

With this manner, the Hamiltonian~(\ref{Eq.33}) turns to be
\begin{equation}
\label{Eq.35}
H_{S}=\frac{p^{2}}{m}-V \sum_{k} \frac{\Delta}{2 E_{k}}\left(C_{k}^{\dagger} C_{-k}^{\dagger} e^{-i k \cdot r}+C_{-k} C_{k} e^{i k \cdot r}\right)+\frac{\Delta^{2}}{V} .\end{equation}

We now introduce the displacement transformation
\begin{equation}
\label{Eq.36}
U_{1}=e^{-i \alpha_{1} \sum_{k} C_{k}^{\dagger} C_{k} k \cdot r},\end{equation}
noting that $U_{1}^{-1} p U_{1}=p-\alpha_{1} \sum_{k} \hbar k C_{k}^{\dagger} C_{k}$ and $U_{1}^{-1}$ $C_{k}^{\dagger}$ $C_{-k}^{\dagger}$ $U_{1}=C_{k}^{\dagger} C_{-k}^{\dagger} e^{i \alpha_{1} k \cdot r}$, we have
\begin{equation}
\label{Eq.37}
\begin{aligned}
\widetilde{H}_{S}&=U_{1}^{-1} H_{S} U_{1}=\frac{p^{2}}{m}(1  \left.-\alpha_{1}\right)^{2}\\
&+\alpha_{1}^{2} \sum_{k} \epsilon_{k}\left(C_{k}^{\dagger} C_{k}+C_{-k}^{\dagger} C_{-k}\right)- \\
& -V \sum_{k} \frac{\Delta}{2 E_{k}}\left[C_{k}^{\dagger} C_{-k}^{\dagger} e^{-i\left(1-\alpha_{1}\right) k \cdot r}+\right. \\
&\left.C_{-k} C_{k} e^{i\left(1-\alpha_{1}\right) k \cdot r}\right]  +\frac{\Delta^{2}}{V},
\end{aligned}
\end{equation}
where $\alpha_1$ is the displaced parameter, for the weak $C_{k\uparrow}-C_{-k\downarrow}$ interaction, $\alpha_1=1$, while for the strong $C_{k\uparrow}-C_{-k\downarrow}$ interaction, $\alpha_1=0$.

Go a step further, noting that the two-mode electrons $C_k$ and $C_{-k}$ are in the mutually local state, they are no longer itinerant electrons, the Cooper pairs are no longer itinerant electron pairs; we use the strong local variational approach to solve the strong coupling superconductivity, with the local variational wave function
\begin{equation}
\label{Eq.38}
\Phi(r)=\left(\frac{\lambda}{\sqrt{\pi}}\right)^{3 / 2} e^{-\frac{1}{2} \lambda^{2} r^{2}} .\end{equation}
Considering $\langle\Phi(r)|e^{i\left(1-\alpha_{1}\right) k \cdot r}| \Phi(r)\rangle$ $=e^{-\left(1-\alpha_{1}\right)^{2} k^{2} / 4 \lambda^{2}}$, and $\langle\Phi(r)\left|\frac{P^{2}}{2 m}\right| \Phi(r)\rangle$ $=\frac{3}{2} \lambda^{2} r_{0}^{2}$ with $r_{0}^{2}$ $=\hbar^{2} / 2 m$, one obtain
\begin{equation}
\label{Eq.39}
\begin{aligned}
\tilde{E}(0)&=\langle\Phi|\widetilde{H}_{S}| \Phi\rangle\\
&=3 \lambda^{2} r_{0}^{2}\left(1-\alpha_{1}\right)^{2}+\frac{\Delta^{2}}{V}\\
&+\alpha_{1}^{2} \sum_{k} \epsilon_{k}\left(C_{k}^{\dagger} C_{k}+C_{-k}^{\dagger} C_{-k}\right)\\
&-V\sum_{k} \frac{\Delta}{2 E_{k}}\left(C_{k}^{\dagger} C_{-k}^{\dagger}+C_{-k} C_{k}\right) e^{-\left(1-\alpha_{1}\right)^{2} k^{2} / 4 \lambda^{2}},
\end{aligned}
\end{equation}
and
\begin{equation}
\label{Eq.40}
\Delta=\frac{1}{2} V \sum_{k}\left(\langle C_{-k} C_{k}\right\rangle+\langle C_{k}^{\dagger} C_{-k}^{\dagger}\rangle) e^{-\left(1-\alpha_{1}\right)^{2} k^{2} / 4 \lambda^{2}} .\end{equation}

At the same time, the attractive two-mode local fluctuation stacking force develops to
\begin{equation}
\label{Eq.41}
\tilde{V}_{\text {st }}=-\frac{9}{4}\left(1-\alpha_{1}\right)^{2} \left(1+\frac{8}{3\pi}\right) \lambda^{2} r_{0}^{2} .\end{equation}

Introducing the new notations
\begin{equation}
\label{Eq.42}
\beta=\frac{1-\alpha_{1}}{\sqrt{2 \lambda_{0}} \alpha_{1}}, \lambda_{0}=2 \lambda^{2} r_{0}^{2},\end{equation}
we can get the following cases:
\begin{enumerate}
\item For the weak coupling, $\alpha_{1} \rightarrow 1$, $\beta \rightarrow 0$, $1-\alpha_{1} \approx \sqrt{2 \lambda_{0}} \beta$, so that $3 \lambda r_{0}^{2}\left(1-\alpha_{1}\right)^{2} \approx 3 \lambda_{0}^{2} \beta^{2}$, and $e^{-\left(1-\alpha_{1}\right)^{2} k^{2} / 4 \lambda^{2}} \approx e^{-\beta^{2} \epsilon_{k}}$.
\item For the strong coupling, $\alpha_{1} \rightarrow 0, \beta \rightarrow \infty, 1-\alpha_{1} \approx 1-\frac{1}{\sqrt{2 \lambda_{0}} \beta}$, so that $3 \lambda r_{0}^{2}\left(1-\alpha_{1}\right)^{2} \approx \frac{3}{2} \lambda_{0}$, and $e^{-\left(1-\alpha_{1}\right)^{2} k^{2} / 4 \lambda^{2}} \approx e^{-\epsilon_{k} / 2 \lambda_{0}}$.
\end{enumerate}

For this paper, we mainly consider the strong coupling superconductivity, thus
\begin{equation}
\label{Eq.43}
\begin{aligned}
\tilde{E}(0)=&\frac{3}{2} \lambda_{0}-\frac{V}{2} \sum_{k} \frac{\Delta}{2 E_{k}}\left[\langle C_{k}^{\dagger} C_{-k}^{\dagger}\rangle+\langle C_{-k} C_{k}\rangle\right]e^{-\epsilon_{k} / 2 \lambda_{0}}\\
+&\frac{\Delta^{2}}{V},
\end{aligned}
\end{equation}
\begin{equation}
\label{Eq.44}
\begin{aligned}
\Delta=\frac{1}{2} V \sum_{k}\left[\langle C_{-k} C_{k}\rangle
+\langle C_{k}^{\dagger} C_{-k}^{\dagger}\rangle\right] e^{-\epsilon_{k} / 2 \lambda_{0}} .
\end{aligned}\end{equation}
\section{Coherent interaction and action-counteraction principles, self-consistent solutions of the energy variational extremum equation and the energy gap equation}
According to the quaternion number method, we have found 
\begin{equation}
\label{Eq.45}
\begin{array}{c}
a^{2}+b^{2}=\cos ^{2} z=\cos z_{+} \cos z_{-},\\
c^{2}+d^{2}=\sin ^{2} z=\sin z_{+} \sin z_{-},
\end{array}\end{equation}
with $\cos z_{+}=a+i b, \cos z_{-}=a-i b$, and $\sin z_{+}=c+i d, \sin z_{-}=c-i d$.

On the other hand, in accordance with the two-mode squeeze transformation, one gets
\begin{equation}
\label{Eq.46}
\begin{array}{c}
C_{k}=\tilde{C}_{k} \cos z_{\mp}+\tilde{C}_{-k}^{\dagger} \sin z_{ \pm}, \\
C_{-k}=\tilde{C}_{-k} \cos z_{\mp}-\tilde{C}_{k}^{\dagger} \sin z_{ \pm}, \\
C_{k}^{\dagger}=\tilde{C}_{k}^{\dagger} \cos z_{\mp}^{*}+\tilde{C}_{-k} \sin z_{ \pm}^{*}, \\
C_{-k}^{\dagger}=\tilde{C}_{-k}^{\dagger} \cos z_{\mp}^{*}-\tilde{C}_{k} \sin z_{ \pm}^{*} .
\end{array}\end{equation}

By mapping to the coherent state representation, we have
\begin{equation}
\label{Eq.47}
\begin{array}{c}
\left|\eta_{+}\right\rangle  =\left|\tilde{\eta}_{+}\right\rangle \cos z_{\mp}+\left|\tilde{\eta}_{-}\right\rangle^{*} \sin z_{ \pm}, \\ \left|\eta_{-}\right\rangle=\left|\tilde{\eta}_{-}\right\rangle \cos z_{\mp}-\left|\tilde{\eta}_{+}\right\rangle^{*} \sin z_{ \pm}, \\
\left|\eta_{+}\right\rangle^{*}  =\left|\tilde{\eta}_{+}\right\rangle^{*} \cos z_{\mp}^{*}+\left|\tilde{\eta}_{-}\right\rangle \sin z_{ \pm}^{*}, \\ \left|\eta_{-}\right\rangle^{*}=\left|\tilde{\eta}_{-}\right\rangle^{*} \cos z_{\mp}^{*}-\left|\tilde{\eta}_{+}\right\rangle \sin z_{ \pm}^{*} .
\end{array}\end{equation}
 
Whereby, based on the coherent interaction representation, we arrive at
%\begin{equation}
%\label{Eq.48}
%\begin{array}{l}\left.\left[\left\langle C_{-k} C_{k}\right\rangle+\left\langle C_{k}^{\dagger} C_{-k}^{\dagger}\right\rangle\right]\right|_{\text {coherent interaction represention }}>0 \\ =\left(\left|\tilde{\eta}_{-}\right\rangle \cos z_{\mp}-\left|\tilde{\eta}_{+}\right\rangle^{*} \sin z_{+}\right)\left(\left|\tilde{\eta}_{+}\right\rangle \cos z_{\mp}+\left|\tilde{\eta}_{-}\right\rangle^{*} \sin z_{ \pm}\right) \\ +\left(\left|\tilde{\eta}_{+}\right\rangle^{*} \cos z_{\mp}^{*}+\left|\tilde{\eta}_{-}\right\rangle \sin z_{ \pm}^{*}\right)\left(\left|\tilde{\eta}_{-}\right\rangle^{*} \cos z_{\mp}^{*}-\left|\tilde{\eta}_{+}\right\rangle \sin z_{ \pm}^{*}\right) \\ =\frac{1}{2}\left\{\left[\left(\cos z_{-} \sin z_{+}+\cos z_{+} \sin z_{-}\right)-\left(\begin{array}{c}\sin z_{+} \cos z_{+} \\ +\sin z_{-} \cos z_{-}\end{array}\right)\right]\right. \\ \left.+\left[\left(\sin z_{+}^{*} \cos z_{-}^{*}+\sin z_{-}^{*} \cos z_{+}^{*}\right)-\left(\begin{array}{c}\cos z_{-}^{*} \sin z_{-}^{*} \\ +\cos z_{+}^{*} \sin z_{+}^{*}\end{array}\right)\right]\right\} \\ =\frac{1}{2}\left\{\left[\begin{array}{c}(a-i b)(c+i d)+(a+i b)(c-i d)- \\ (c+i d)(a+i b)-(c-i d)(a-i b)\end{array}\right]\right. \\ \left.+\left[\begin{array}{c}(c-i d)(a+i b)+(c+i d)(a-i b)- \\ (a+i b)(c+i d)-(a-i b)(c-i d)\end{array}\right]\right\} \\ =2\left(\frac{\Delta}{2 E_{k}}+\frac{\Delta}{2 E_{k}} \cdot \frac{\epsilon_{k}}{E_{k}}\right).
%\end{array}\end{equation}
\begin{equation}
\begin{split}
&[\langle C_{-k}C_k \rangle + \langle C_k^{\dagger}C_{-k}^{\dagger} \rangle|_{\text{coherent interaction represention}}>0\\
=&(|\tilde{\eta}_{-}\rangle \cos z_{\mp}-|\tilde{\eta}_{+}\rangle^{*} \sin z_{\pm})(|\tilde{\eta}_{+}\rangle \cos z_{\mp} + |\tilde{\eta}_{-}\rangle^{*} \sin z_{\pm})_{\text{max}}\\
+&(|\tilde{\eta}_{+}\rangle^{*} \cos z_{\mp}^{*}+|\tilde{\eta}_{-}\rangle \sin z_{\pm}^{*})(|\tilde{\eta}_{-}\rangle^{*} \cos z_{\mp}^{*}-|\tilde{\eta}_{+}\rangle \sin z_{\pm}^{*})_{\text{max}}\\
=&\frac{1}{2}\{(\cos z_{-} \sin z_{+}+\cos z_{+} \sin z_{-})-(\sin z_{+}\cos z_{+}+\sin z_{-} \cos z_{-})\\
&+(\sin z_{+}^{*} \cos z_{-}^{*}+\sin z_{-}^{*} \cos z_{+}^{*})-(\cos z_{-}^{*} \sin z_{-}^{*}+\cos z_{+}^{*} \sin z_{+}^{*})\}\\
=&\frac{1}{2}\{(a-ib)(c+id)+(a+ib)(c-id)-(c+id)(a+ib)\\
&-(c-id)(a-ib)+(c-id)(a+ib)+(c+id)(a-ib)\\
&-(a+ib)(c+id)-(a-ib)(c-id)\}\\
=&2\left(\frac{\Delta}{2E_k}+\frac{\Delta}{2E_k} \times \frac{\epsilon_k}{E_k}\right).
\end{split}
\end{equation}

Therefore, by virtue of the coherent interaction between the one-electron coherent states $\left|\eta_{ \pm}\right\rangle$and the two-mode coherent state $\left|z_{\uparrow \downarrow}\right\rangle$, we find for $\tilde{E}(0)$ and $\Delta$ as
\begin{equation}
\label{Eq.49}
\tilde{E}(0)=\frac{3}{2} \lambda_{0}-V \sum_{k} \frac{\Delta}{2 E_{k}} \frac{\Delta}{2 E_{k}}\left(1+\frac{\epsilon_{k}}{E_{k}}\right) e^{-\epsilon_{k} / 2 \lambda_{0}}, 
\end{equation}
\begin{equation}
\label{Eq.50}
\Delta=V \sum_{k} \frac{\Delta}{2 E_{k}}\left(1+\frac{\epsilon_{k}}{E_{k}}\right) e^{-\epsilon_{k} / 2 \lambda_{0}} .
\end{equation}

For the superconducting Hamiltonian, since the one-mode electrons $C_{k \uparrow}, C_{-k \downarrow}$ and the two-mode Cooper pair $C_{-k \downarrow} C_{k \uparrow}$ are to occur simultaneously, the one-electron coherent state $\left|\eta_{ \pm}\right\rangle$and the two-mode squeezed state $\left|z_{\uparrow \downarrow}\right\rangle$ are also to occur at the same time. As well known, the squeezed state action is a two-order nonlinear quantum fluctuation effect, as a result the coherent state $\left|\eta_{ \pm}\right\rangle$will evolve to the squeezed coherent state $\left|\tilde{\eta}_{ \pm}\right\rangle$. In fact, using Eq.~(\ref{Eq.25}) and Eq.~(\ref{Eq.31}) one can obtain
\begin{equation}
\label{Eq.51}
\begin{split}
&e^{C_{k}^{\dagger} \eta_{+}} s(z) C_{k} s^{\dagger}(z) e^{-C_{k}^{\dagger} \eta_{+}}\\
=&\left(C_{k} \cos z-C_{-k}^{\dagger} \sin z\right)-\left(2 s_{+} \eta_{+} \cos z-2 s_{-} \eta_{-}^{*} \sin z\right), \\
&e^{\eta_{+}^{*} C_{k}} s(z) C_{k}^{\dagger} s{\dagger}(z) e^{-\eta_{+}^{*} C_{k}}\\
=&\left(C_{k}^{\dagger} \cos z-C_{-k} \sin z\right)-\left(2 s_{+} \eta_{+}^{*} \cos z-2 s_{-} \eta_{-} \sin z\right), \\
&e^{C_{-k}^{\dagger} \eta_{-}} s(z) C_{-k} s{\dagger}(z) e^{-C_{-k}^{\dagger} \eta_{-}}\\
=&\left(C_{-k} \cos z+C_{k}^{\dagger} \sin z\right)-\left(2 s_{-} \eta_{-} \cos z+2 s_{+} \eta_{+}^{*} \sin z\right), \\
&e^{\eta_{-}^{*} C_{-k}} S(z) C_{-k}^{\dagger} s{\dagger}(z) e^{-\eta_{-}^{*} C_{-k}}\\
=&\left(C_{-k}^{\dagger} \cos z+C_{k} \sin z\right)-\left(2 s_{-} \eta_{-}^{*} \cos z+2 s_{+} \eta_{+} \sin z\right).
\end{split}
\end{equation}

From above results of Eq.~(\ref{Eq.51}), we can find the evolutions for the coherent parameters $\eta_{+}$and $\eta_{-}$
\begin{equation}
\label{Eq.52}
\begin{split}
\eta_{+} \rightarrow \tilde{\eta}_{+}&=\langle 0\left|2 s_{+} \cos z+2 s_{+} \sin z\right| 0\rangle \eta_{+}\\
&=(\cos z+\sin z) \eta_{+},\\
\eta_{-} \rightarrow \tilde{\eta}_{-}&=\langle 0\left|2 s_{-} \cos z-2 s_{-} \sin z\right| 0\rangle \eta_{-}\\
&=(\cos z-\sin z) \eta_{-}.
\end{split}
\end{equation}

That is, the coherent state $\left|\eta_{+}\right\rangle$ evolves to the squeezed coherent state $\left|\tilde{\eta}_{+}\right\rangle=e^{C_{k \uparrow}^{\dagger}(\cos z+\sin z) \eta_{+}}|0\rangle$, while the coherent state $\left|\eta_{-}\right\rangle$evolves to $\left|\tilde{\eta}_{-}\right\rangle=e^{C_{-k \downarrow}^{\dagger}(\cos z-\sin z) \eta_{-}}|0\rangle$. Accordance with the action-counteraction principles of natural philosophy, the two-mode squeezed state $\left|z_{\uparrow \downarrow}\right\rangle$ also evolves to $\left|\tilde{z}_{\uparrow \downarrow}\right\rangle=e^{\sum_{\mathrm{k}}\left(\tilde{z} C_{k \uparrow}^{\dagger} C_{-k \downarrow}^{\dagger}-\tilde{z}^{*} C_{-k \downarrow} C_{k \uparrow}\right)}|0\rangle ~$ with $~ \tilde{z}=(\cos z+$ $\sin z) z_{\uparrow \downarrow}+(\cos z-\sin z) z_{\uparrow \downarrow}$. As above stated, the two-mode squeezed state $\left|z_{\uparrow \downarrow}\right\rangle$ is originated from the two-mode attractive action $-V \sum\left(C_{k \uparrow}^{\dagger} C_{-k \downarrow}^{\dagger}+C_{-k \downarrow} C_{k \uparrow}\right)$, on account of the action-counteraction effect, $V$ will be corrected as
\begin{equation}
\label{Eq.54}
\begin{aligned}
\tilde{V}&=(\cos z+\sin z) V  +(\cos z-\sin z) V \\
& =\sqrt{2}\left(1+\frac{\epsilon_{k}}{E_{k}}\right)^{\frac{1}{2}} V .
\end{aligned}\end{equation}

Finally, $\tilde{E}(0)$ and $\Delta$ turn out to be
\begin{equation}
\label{Eq.55}
\begin{aligned}
\tilde{E}(0)&=\frac{3}{2} \lambda_{0}-\sqrt{2} V \sum_{k}\left(\frac{\Delta}{2 E_{k}}\right)^{2}\left(1+\frac{\epsilon_{k}}{E_{k}}\right)^{\frac{3}{2}}
e^{-\epsilon_{k} / 2 \lambda_{0}}\\
&+\frac{\Delta^{2}}{V}, 
\end{aligned}\end{equation}
\begin{equation}
\label{Eq.56}
\begin{aligned}
\Delta=\sqrt{2} V \sum_{k} \frac{\Delta}{2 E_{k}}\left(1+\frac{\epsilon_{k}}{E_{k}}\right)^{\frac{3}{2}} e^{-\epsilon_{k} / 2 \lambda_{0}} .
\end{aligned}\end{equation}

Further, transforming $\sum_{k}(\cdots)$ to integral $\int(\cdots) d \epsilon$ with the state density $g(0)$ and $\hbar \omega_{0}=2 \pi \Delta_{0}(\xi)$, we have for $\tilde{E}(0)$
\begin{equation}
\label{Eq.57}
\begin{aligned}
\tilde{E}(0)&=\frac{3}{2} \lambda_{0}-\sqrt{2} V g(0) \int_{0}^{\hbar \omega_{0}} \frac{d \epsilon}{2\left(\epsilon^{2}+\Delta^{2}\right)} \Delta^{2}\\
&\times \left(1+\frac{\epsilon}{\sqrt{\epsilon^{2}+\Delta^{2}}}\right)^{\frac{3}{2}} e^{-\epsilon / 2 \lambda_{0}}+\frac{\Delta^{2}}{V}.
\end{aligned}\end{equation}

From Eq.~(\ref{Eq.57}) we obtain the variational extremum equation
\begin{equation}
\begin{array}{c}
\label{Eq.58}
\frac{\partial\tilde{E}(0)}{\partial\lambda_{0}}=0,\\
\frac{3}{2}-\frac{\sqrt{2}Vg(0)}{4\lambda_{0}^{2}}\int_{0}^{\hbar\omega_{0}(\xi)}\frac{\epsilon d\epsilon}{\left(\epsilon^{2}+\Delta^{2}\right)}\Delta^{2}\left(1+\frac{\epsilon}{\sqrt{\epsilon^{2}+\Delta^{2}}}\right)^{\frac{3}{2}}e^{-\epsilon/2\lambda_{0}}=0,
\end{array}
\end{equation}
that is
\begin{equation}
\label{Eq.59}
\begin{split}
   \frac{3}{2}&=\frac{\sqrt{2} V g(0)}{4 \lambda_{0}^{2}} \int_{0}^{\hbar \omega_{0}(\xi)} \frac{\epsilon d \epsilon}{\left(\epsilon^{2}+\Delta^{2}\right)} \Delta^{2}\\
&\times \left(1+\frac{\epsilon}{\sqrt{\epsilon^{2}+\Delta^{2}}}\right)^{\frac{3}{2}} e^{-\epsilon / 2 \lambda_{0}}, 
\end{split}
\end{equation}
while the energy gap equation
\begin{equation}
\begin{split}
    1&=\sqrt{2} V g(0) \int_{0}^{\hbar \omega_{0}(\xi)} \frac{d \epsilon}{\sqrt{\epsilon^{2}+\Delta^{2}}}\\
&\times \left(1+\frac{\epsilon}{\sqrt{\epsilon^{2}+\Delta^{2}}}\right)^{\frac{3}{2}} e^{-\epsilon / 2 \lambda_{0}}.
\end{split}
\label{Eq.60}
\end{equation}

(i) Weak coupling

Obviously, for the weak coupling, $\alpha_{1} \rightarrow 1, \beta \rightarrow 0$, it returns to the BCS Hamiltonian, while the energy-gap equation is
\begin{equation}
\label{Eq.61}
1=\sqrt{2} V g(0) \int_{0}^{\hbar \omega_{0}} \frac{d \epsilon}{\sqrt{\epsilon^{2}+\Delta^{2}}}\left(1+\frac{\epsilon}{\sqrt{\epsilon^{2}+\Delta^{2}}}\right)^{\frac{3}{2}} .\end{equation}

The numerical results for the BCS Hamiltonian are shown in Tab.~(\ref{Tab1}).

\begin{table}[ht]
\caption{The superconducting energy gap $\Delta_{\mathrm{A \cdot C}}$ associated with the coherent interaction and action-counteraction, in which the weak coupling strength $V g(0)=0.1, 0.2, 0.3$ and $0.4$ , with $\hbar \omega_{0}=0.01~\text{eV} $.}
\label{Tab1}
\centering
\begin{ruledtabular}
\begin{tabular}{cccc}
\textrm{$V g(0)$} &
\textrm{$\Delta_{\text{BCS}} / \mathrm{eV}$} &
%\multicolumn{1}{c}{$\Delta_{A \cdot C} / \mathrm{eV}$} &
\textrm{$\Delta_{\mathrm{A \cdot C}} / \mathrm{eV}$} \\
\colrule
0.1 & $9.08 \times 10^{-6}$ & 0.000978 \\
0.2 & $1.35 \times 10^{-4}$ & 0.00372 \\
0.3 & $7.14 \times 10^{-4}$ & 0.00618 \\
0.4 & $1.65 \times 10^{-3}$ & 0.00828 \\
\end{tabular}
\end{ruledtabular}
\end{table}

From Tab.~(\ref{Tab1}) we see that, for the weak coupling, the principles of the coherent interaction and action-counteraction noticeably increases the superconducting energy gap, for $Vg(0)=0.1, \Delta_{\mathrm{A \cdot C}} \approx 108 \Delta_{\text{BCS}}$, but for $Vg(0)=0.2, \Delta_{\mathrm{A \cdot C}} \approx 28 \Delta_{\text{BCS}}$.

(ii) Strong coupling

For the strong coupling, $\alpha_{1} \rightarrow 0, \beta \rightarrow \infty$, as a result, $1-\alpha_{1} \approx 1-\frac{1}{\sqrt{2 \lambda_{0} \beta}}$, as well shown in Eq.~(\ref{Eq.59}) and Eq.~(\ref{Eq.60}). As it is known, the cut-off frequency $\hbar \omega_{0}(\xi)$ depends on the coherent length $\xi$ and the energy-gap $\Delta$, that is to say, $\hbar \omega_{0}(\xi)$ also depend upon coupling strength $V g(0)$ because the energy-gap varies with the coupling strength. From Eq.~(\ref{Eq.59}) and Eq.~(\ref{Eq.60}), for a given $\hbar \omega_{0}(j)$, one can look for $\lambda_{0}(j)$ and $\Delta(j)$ by the self-consistent manner; in addition, $\lambda_{0}$ and $\Delta$ increase with increasing $\hbar \omega_{0}(\xi)=2 \pi \Delta_{0}(\xi)$. On the other hand, because Eq.~(\ref{Eq.59}) and Eq.~(\ref{Eq.60}) are restricted each other for a given $\Delta_{0}$, the integral value of Eq.~(\ref{Eq.59}) will decrease with $\left(\lambda_{0} \uparrow, \Delta \downarrow\right)$, but the integral value of Eq.~(\ref{Eq.60}) will increase; on the contrary, the integral value of Eq.~(\ref{Eq.59}) will increase with $\left(\lambda_{0} \downarrow, \Delta \uparrow\right)$, but the integral value of Eq.~(\ref{Eq.60}) will decrease. Therefore, considering the variations of $\left\{\lambda_{0}\left(j_{i}\right), \Delta\left(j_{i}\right)\right\}$ with $\Delta_{0}\left(j_{i}\right), j_{i}=j_{1}, j_{2}, \cdots$, one will encounter that when $\Delta_{0}\left(j_{i}\right)$ evolves to $\Delta_{0}\left(j_{\text{c}}\right)$, the $\Delta$ reaches its maximum $\Delta\left(j_{\text{c}}\right)$ at $\lambda_{0}\left(j_{\text{c}}\right)$, i.e., $\Delta\left(j_{\text{c}}\right)=\Delta_{\mathrm{A \cdot C}}$, as shown in Tab.~(\ref{Tab2}) and Tab.~(\ref{Tab3}). From Tab.~(\ref{Tab2}) we can see that, for the strong coupling $\alpha_{1}=0$, when $Vg(0)$ is given, the $e^{-\frac{k^{2}}{4 \lambda^{2}}}=e^{-\frac{\epsilon}{2 \lambda_{0}}}=$ const from beginning to end with $\epsilon$ since $\frac{k^{2}}{\lambda^{2}}=$ const. Therefore, for the self-consistent equations Eq.~(\ref{Eq.59}) and Eq.~(\ref{Eq.60}), the unknown parameters $\hbar \omega_{0}(\xi)$ and $\Delta$ will be existed as the unique solution. That is, we will find the integral upper limit $\hbar \omega_{0}(\xi)$, while $\Delta$ also has its certain value at the same time.
\begin{table}[ht]
\caption{Example: for $V g(0)=1.4, \Delta(j)$ increases with $\Delta_{0}(j)$ increasing, and finally reaches to a maximum $\Delta=0.5676~\text{eV} $ at $\Delta_{0}=0.0780~\text{eV}, \lambda_{0}=0.1270~\text{eV}$; but the reduced coupling strength $\tilde{V} g(0) \approx 0.2$ from beginning to end with $\epsilon_{j}$.}
\label{Tab2}
\centering
\begin{ruledtabular}
\begin{tabular}{ccccccc}
\textrm{$\epsilon_{j}/\text{eV}$} &
\textrm{$\Delta_{0}(j)/\text{eV} $} &
\textrm{$\lambda_{0}(j)/\text{eV} $} &
\textrm{$\Delta(j)/\text{eV} $} &
\textrm{$\epsilon_{j}/2 \lambda_{0}$} &
\textrm{$e^{-\epsilon_{j}/2 \lambda_{0}}$} &
\textrm{$\tilde{V}g(0)$} \\
\colrule
0.0100 & 0.0016 & 0.0026 & 0.0120 & 1.95 & 0.142 & 0.199 \\
0.0812 & 0.0129 & 0.0208 & 0.0960 & 1.95 & 0.142 & 0.199 \\
0.1724 & 0.0274 & 0.0442 & 0.1960 & 1.95 & 0.142 & 0.199 \\
0.2592 & 0.0413 & 0.0665 & 0.2800 & 1.95 & 0.142 & 0.199 \\
0.3456 & 0.0550 & 0.0890 & 0.3800 & 1.94 & 0.144 & 0.201 \\
0.4147 & 0.0660 & 0.1070 & 0.4680 & 1.94 & 0.144 & 0.201 \\
0.4900 & 0.0780 & 0.1270 & 0.5676 & 1.94 & 0.144 & 0.201
\end{tabular}
\end{ruledtabular}
\end{table}

\begin{table}[ht]
\caption{Based on the strong coupling local approach, the superconducting energy gap $\Delta_{\mathrm{A \cdot C}}$, associated with the coherent interaction and action-counteraction, versus the coupling strength $Vg(0)= 1.2,1.4,1.6$ and $1.8$.}
\label{Tab3}
\centering
\begin{center}
Eq.~(\ref{Eq.60})
\end{center}
\begin{ruledtabular}
\begin{tabular}{cccccc}
\textrm{$Vg(0)$} &
\textrm{$\Delta_{0}/\text{eV}$} &
\textrm{$\lambda_{0}(j)/\text{eV}$} &
\textrm{$\Delta_{\mathrm{A \cdot C}}/\text{eV}$} &
\textrm{1.5} &
\textrm{$\tilde{V}_{\text{st}}/\text{eV}$} \\
\colrule
1.2 & 0.0658 & 0.085 & 0.3480 & 1.501 & 0.177 \\
1.4 & 0.0780 & 0.127 & 0.5676 & 1.500 & 0.264 \\
1.6 & 0.0950 & 0.178 & 0.8620 & 1.501 & 0.370 \\
1.8 & 0.1052 & 0.219 & 1.0850 & 1.501 & 0.456 \\
\end{tabular}
\end{ruledtabular}
\begin{center}
Eq.~(\ref{Eq.61})
\end{center}
\begin{ruledtabular}
\begin{tabular}{cccccc}
\textrm{$Vg(0)$} &
\textrm{$\Delta_{0}/\text{eV}$} &
\textrm{$\lambda_{0}(j)/\text{eV}$} &
\textrm{$\Delta_{\mathrm{A \cdot C}}/\text{eV}$} &
\textrm{1.0} &
\textrm{$\tilde{V}_{\text{st}}/\text{eV}$} \\
\colrule
1.2 & 0.0658 & 0.085 & 0.3480 & 1.034 & 0.177 \\
1.4 & 0.0780 & 0.127 & 0.5676 & 1.010 & 0.264 \\
1.6 & 0.0950 & 0.178 & 0.8620 & 0.995 & 0.370 \\
1.8 & 0.1052 & 0.219 & 1.0850 & 1.033 & 0.456 \\
\end{tabular}
\end{ruledtabular}
\end{table}

For the coupling strength $V g(0)=1.4$, it will reduce to the weak coupling $\tilde{V} g(0) \approx 0.2$ from beginning to end with $\epsilon$, while the Cooper pair reduces to $\widetilde{C_{k \uparrow} C_{-k \downarrow}}=0.14 C_{k \uparrow} C_{-k \downarrow}$, the small Cooper pair. In virtue of this result, the coherent interaction and action-counteraction will play a more remarkable function, $\Delta_{\mathrm{A \cdot C}}=0.5676~\text{eV} \gg \hbar \omega_{\text{D}} \gg \Delta_{\text{BCS}}$, as shown in Tab.~(\ref{Tab3}). Importantly, based on the strong local fluctuation action, $\tilde{V}_{\text{st}}=0.264~\text{eV}  \gg \hbar \omega_{\text{D}} (\sim0.01~\text{eV} )$
\section{Conclusions and discussion}
In synthesizing the results in sections 2-4, some key physics and principal comments are obtained.

(1) For the Watson's two helix model of DNA, the base pairing with the stacking force but not with the hydrogen bond is the biology principles of natural philosophy. Equally, for the superconducting mechanism, the two-mode $C_{k \uparrow}C_{-k \downarrow}$ pairing with the strong local fluctuation stacking force but not with the attractive electron-election interaction $V_{\text{el-el}}(\text{e-p})$ (or another mechanism) is the physical principles of natural philosophy, in which the attractive strongly local fluctuation stacking force $\tilde{V}_{\text{st}}=-\frac{9}{4} \left(1-\alpha_{1}\right)^{2}\left(1+\frac{8}{3\pi}\right)\lambda^{2} r_{0}^{2}$. Since the tying Cooper pairing $C_{-k \downarrow}C_{k \uparrow}e^{ik\cdot r}$ replaces the itinerant Cooper pairing $C_{-k \downarrow}C_{k \uparrow}$, it is not possible to exchange the original pairing to make a new pairing in the superconductors , that is the most important new idea to greatly increase the ground state energy gap for solving high-Tc superconductivity. 

At the initial stage, since $V_{\text{el-el}}(\text{e-p})$ leads to the opposite direction motion of the electrons $C_{k \uparrow}$ and $C_{-k \downarrow}$ with acceleration $a_{1}$, with the time evolution $k$ will develop to a larger value due to $p=\hbar k$. Noting that $\langle\Phi(r)|\frac{p^{2}}{m}| \Phi(r)\rangle=3 \lambda^{2} r_{0}^{2}$ and $p^{2}=k^{2} \hbar^{2}$, this shows $\frac{k^{2}}{\lambda^{2}}=$ const. That is to say, with the time evolution, $\lambda^{2}$ will also develop to a larger value. At the same time, the local fluctuation stacking force also further promotes the opposite direction motion with acceleration $a_{2} \gg a_{1}$. With this evolution result, $a_{2}$ develops to a larger value, at the end, the attractive local fluctuation stacking force $\tilde{V}_{\text{st}}$ towards a very large value, $\tilde{V}_{\text{st}} \gg \hbar \omega_{\text{D}}$. For the strong coupling $\alpha_{1}=0, \tilde{V}_{\text{st}} \gg \hbar \omega_{\text{D}}$, while for the weak couple $\alpha_{1}=1, \tilde{V}_{\text{st}}=0$, that is the physical principles of natural philosophy for superconducting mechanism. For the coupling strength $V g(0)=1.4, \tilde{V}_{\text{st}}=0.264~\text{eV} \gg \hbar \omega_{\text{D}}$, with $\lambda_{0}(c)=0.127~\text{eV} $ (as sec in Tab.~(\ref{Tab2})).%Here!

(2) Quantum fluctuation effect is a very important quantum effect in physics. For the two-mode electron attractive interaction of superconductors, the strong local quantum fluctuation action are far beyond the existing fluctuation effects, such as resonating valence bond fluctuations,charge fluctuations, spin fluctuations, and antiferromagnetism fluctuations. From the strong local idea, $\langle\Delta^{2} x\rangle|_{\lambda \rightarrow \infty} \rightarrow 0$ the two-mode electrons $C_{k\uparrow}$ and $C_{-k\downarrow}$ are in a strongly local stacking state, the attractive interaction (when $\alpha_1=0$)
\begin{equation}
\label{Eq.62}
\begin{split}
V_{\text{st}}=&\langle \Phi|\frac{p^2}{2m}+\frac{p^2}{2m}|\Phi\rangle s_1(\uparrow) \cdot s_2(\downarrow)\\
=&-\frac{9}{4}\left(1-\alpha_1^2\right)^2 \lambda^2 r_0^2 \gg \hbar \omega_{\text{D}}.
\end{split}
\end{equation}
However, since $p=-i \hbar \nabla, \langle\Phi |p|\Phi \rangle=i\hbar \frac{2}{\sqrt{\pi}} \lambda$, by considering the local quantum fluctuation action $\langle \Phi| \Delta^2 p| \Phi \rangle=\langle \Phi |p^2| \Phi \rangle-\langle \Phi|p| \Phi \rangle^2$,
\begin{equation}
\label{Eq.63}
\begin{split}
\tilde{V}_{\text{st}}=-\frac{9}{4}\left(1+\frac{8}{3\pi}\right)\lambda^2 r_0^2>V_{\text{st}} \gg \hbar \omega_{\text{D}}.
\end{split}
\end{equation}
Based on the principles of coherent interaction and action-counteraction, in view of the strong local variational approach for superconductivity theory with the strong local fluctuation stacking force pairing mechanism, two results can be obtained: 1) Weak coupling, $\beta \rightarrow 0$, with $\alpha_1 \rightarrow 1$ (the two electron parallel orbits are away from each other) and $\lambda_0 \rightarrow 0$ (weak local), i.e. the low-$T_c$ superconductivity; 2) Strong coupling, $\beta \rightarrow \infty$, with $\alpha_1 \rightarrow 0$ (the two electron parallel orbits are very close to each other) and $\lambda_0>0.085$ eV (strong local), i.e. the high-$T_c$ superconductivity and the room-temperature superconductivity.

(3) As we stated in Sec. 3, owing to the two-mode coherent action $H_{\text{c}}^{(2)} = {i\hbar }\sum_{k}\left( \zeta C_{k \uparrow}^{\dagger}C_{- k \downarrow}^{\dagger} - \zeta^{*}C_{- k \downarrow}C_{k \uparrow} \right)$, it will result in the two-mode squeezed state $\left| {{{z}}_{\uparrow \downarrow }} \right\rangle =s\left( z \right)\left| {{\Phi }^{\left( 2 \right)}}\left( 0 \right) \right\rangle$, while the one-electron coherent action ${i\hbar }\sum_{k}\left( f C_{k}^{\dagger} - f C_{k} \right)\ $ will also result in the one-electron coherent states $\left| \eta_{\pm} \right\rangle = D\left( \eta_{\pm} \right)|0\rangle_{c}$. Since they are occurred at the same time, in virtue of the nonlinear two-mode squeezed action $H_{c}^{\left( 2 \right)}$, $\left| \eta_{+} \right\rangle$ will evolve to the squeezed coherent state $\left| {\tilde{\eta}}_{+} \right\rangle = e^{C_{k \uparrow}^{\dagger}(\cos z + \sin z)}|0\rangle$, while $\left| \eta_{-} \right\rangle$ to $\left| {\tilde{\eta}}_{-} \right\rangle = e^{C_{- k \downarrow}^{\dagger}(\cos z - \sin z)}\left| \eta_{-} \right\rangle$. Accordance with the action-counteraction principle, in the meantime, the two-mode squeezed state$\left| z_{\uparrow \downarrow} \right\rangle$also evolves to $\left| {\tilde{z}}_{\uparrow \downarrow} \right\rangle = e^{\sum_{k}^{}{{\tilde{z}}_{\uparrow \downarrow}\left( C_{k \uparrow}^{\dagger}C_{- k \downarrow}^{\dagger} - C_{- k \downarrow}C_{k \uparrow} \right)}}|\Phi^{\left( 2 \right)}(0)\rangle_{\text{c}}$ with ${\tilde{z}}_{\uparrow \downarrow} = \lbrack(\cos z + \sin z) + (\cos z - \sin z)\rbrack z_{\uparrow \downarrow}$. From this effect of the coherent interaction and action-counteraction, the superconducting energy gap is corrected as $\Delta = \sqrt{2}V\sum_{k}\frac{\Delta}{{2E}_{k}}\left( 1 + \frac{\epsilon_{k}}{E_{k}} \right)^{\frac{3}{2}}$, but for the BCS theory, $\Delta = V\sum\frac{\Delta}{2E_{k}}$. That is shown, for the superconducting state, the principal quantum effects are originated from the coherent interaction and action-counteraction and the strong local quantum fluctuations.

(4) As we see in Tab.~(\ref{Tab1}), for $Vg(0) = 0.1,\Delta_{\mathrm{A \cdot C}} \approx 108\Delta_{\text{BCS}}$, however, for ${Vg}(0) = 0.2$, $\Delta_{\mathrm{A \cdot C}} \approx 28.0\Delta_{\text{BCS}}$, and for $Vg(0) = 0.3,\Delta_{\mathrm{A \cdot C}} \approx 8.65\Delta_{\text{BCS}}$. That is about it, the coherent interaction and action-counteraction play a remarkable action only to the weak coupling, but not to the strong coupling. To solve this knotty problem, as a new important idea, we take the tying Cooper pair $C_{k \uparrow}C_{- k \downarrow}e^{{ik} \cdot r}$ to replace the itinerant Cooper pair $C_{k \uparrow}C_{- k \downarrow}$. Go a step further, based on the strong local variational approach $\langle\Phi(r)|C_{k \uparrow}C_{- k \downarrow}e^{ik \cdot r}|\Phi(r)\rangle$ with $\Phi(r) = \left( \frac{\lambda}{\sqrt{\pi}} \right)^{3/2}e^{- \frac{1}{2}\lambda^{2}r^{2}}$, the Cooper pairing transfers to the strong local stacking pairing $C_{k \uparrow}C_{- k \downarrow}e^{- \left( 1 - \alpha_{1} \right)^{2}k^{2}/4\lambda^{2}}\ $ with $\alpha_{1} = 0$. In terms of the equivalent principle, $C_{k \uparrow}C_{- k \downarrow}:~Vg(0) = C_{k \uparrow}C_{- k \downarrow}e^{- k^{2}/4\lambda^{2}}:\tilde{V}g(0)$, the reduced coupling strength $\tilde{V}g(0) = e^{- \frac{k^{2}}{4\lambda^{2}}}{Vg}(0) = e^{-\epsilon/2\lambda_{0}}{Vg}(0)$. For example, if ${Vg}(0) = 1.4$, $\tilde{V}g(0) = e^{- 0.3456/2 \times 0.0890}$ $Vg(0) \approx 0.2$ from beginning to end with $\epsilon$ due to $k^{2}/4\lambda^{2} =$ const, as shown in Tab.~(\ref{Tab2}). At the same time,$C_{k \uparrow}C_{- k \downarrow}e^{- \left( 1 - \alpha_{1} \right)^{2}k^{2}/4\lambda^{2}} \approx 0.14C_{k \uparrow}C_{- k \downarrow}$, the BCS Cooper pair transfers to a small Cooper pair. On the other hand, for the weak coupling, $\alpha_{1} = 1,C_{k \uparrow}C_{- k \downarrow}e^{{- \left( 1 - \alpha_{1} \right)}^{2}k^{2}/4\lambda^{2}} = C_{k \uparrow}C_{- k \downarrow}$, the BCS Cooper pair. By virtue of these new ideas the coherent interaction and action- counteraction still play a remarkable function.For the high -temperature or room -temperature superconductivity theory, two puzzled problems must be solved: (a) the attractive interaction between electrons $C_{k \uparrow}$ and $C_{- k \downarrow}{\ V}_{\text{el-el}} \gg \hbar\omega_{\text{D}}$; (b)the ground state energy gap of superconductivity $\Delta \left( 0 \right)>0.348~\text{eV}\gg \hbar {{\omega }_{\text{D}}}$. Based on the principles of coherent interaction and action-counteraction and the strong local variational approach, importantly, the pairing mechanism of strong local fluctuation force with tying Cooper pair, we have been solved above two puzzled problems successfully. For the strong coupling, $\Delta_{\mathrm{A \cdot C}} \gg \hbar\omega_{\text{D}}$, $\tilde{V}_{\text{st}}\gg \hbar \omega_{\text{D}}$, and $\hbar \omega_0 \gg \hbar \omega_{\text{D}}$. As shown in the following:

\begin{ruledtabular}
	\begin{tabular}{cccccc}
		\textrm{$Vg(0)$} &
		\textrm{$\Delta_{\mathrm{A \cdot C}}/\text{V}$} &
		\textrm{$\tilde{V}_{\text{st}}/\text{eV}$} &
		\textrm{$\hbar \omega_0/V$} \\
		\colrule
		1.2 & 0.3480 & 0.177 & 0.413 \\
		1.4 & 0.5676 & 0.264 & 0.490 \\
		1.6 & 0.8620 & 0.370 & 0.596 \\
		1.8 & 1.0850 & 0.456 & 0.661 \\
	\end{tabular}
\end{ruledtabular}

(5) The integral upper limit of energy $\hbar\omega_{0}(\text{c})( = 2\pi\Delta_{0}(\text{c}))$ shows that the electrons for $\epsilon_{k}>\hbar\omega_{0}(\text{c})$ are fewer in number. For the weak coupling $Vg(0) = 0.1 - 0.3,$ $\Delta \approx 0.0001~\text{eV}  \ll \hbar \omega_{\text{D}}( \approx 0.01~\text{eV} )$, the BCS theory is correct. However, for the strong coupling, $\Delta \gg \hbar\omega_{\text{D}}$, the integral upper limit $\hbar\omega_{\text{D}} = 0.01~\text{eV} $ would eliminate the contribution of a large number of electrons to the superconducting state. Clearly, the integral upper limit $\hbar \omega_{0}(\text{c})$ is a very difficult to solve. This paper the energy extremum equation Eq.~(\ref{Eq.59}) and the energy gap equation Eq.~(\ref{Eq.60}) are restricted each other: the integral value of Eq.~(\ref{Eq.59}) increases with $\left( \lambda_{0} \downarrow \Delta \uparrow \right)$ and decreases with $\left( \lambda_{0} \uparrow \Delta \downarrow \right)$, but on the contrary, the integral value of Eq.~(\ref{Eq.60}) increases with $\left( \lambda_{0} \uparrow \Delta \downarrow \right)$and decreases with $\left( \lambda_{0} \downarrow \Delta \uparrow \right)$. For the strong coupling, when $Vg(0)$ is given, $e^{- \frac{k^{2}}{4\lambda^{2}}} = e^{- \frac{\epsilon}{2\lambda_{0}}} =$ const from beginning to end with $\epsilon$ since $k^{2}/\lambda^{2} =$ const. As this result, the integral upper limit $\hbar\omega_{0}(\text{c})$ and the energy gap $\Delta$ will be given as the unique solution from Eq.~(\ref{Eq.59}) and Eq.~(\ref{Eq.60}), as shown in Tab.~(\ref{Tab2}).

(6) In particular, we would point out that for the current theory of superconductivity, the two-mode attraction $V_{C_k-C_{-k}}\leq V_0 \approx \hbar \omega_{\text{D}}$, for this reason, the resulting energy gap $\Delta \ll \Delta (0) \approx \hbar \omega_{\text{D}}$. Therefore, high-$T_c$ or room-$T_c$ superconductivity is difficult to achieve. For our new theory of superconductivity, based on the strong local quantum fluctuation, since $K^2\uparrow \rightarrow \lambda^2 \uparrow \rightarrow \tilde{V}_{\text{st}}\uparrow \rightarrow K^2\uparrow \rightarrow \lambda^2 \uparrow \rightarrow \tilde{V}_{\text{st}}\uparrow \cdots$, that is to say the development of $\tilde{V}_{\text{st}}$ has no upper limit; namely, since $\Delta=\tilde{V}_{\text{st}}\sum_k \langle C_{-k}C_k \rangle$, so then the development of $\Delta$ has also no upper limit. For a certain superconductor we have a certain energy integral upper limit \( \epsilon_{k} (c) = \frac{\hbar^{2} K_{c}^{2}}{2m} \), accordingly, we determine \( \lambda (c) \) and \( \tilde{V}_{\text{st}} (c) \). $\epsilon_k(c)$ increases as the coupling strength $\tilde{V}_{st}g(0)$ increases, with $\tilde{V}_{st}g(0)\gg 1.0$, we will see $\epsilon_k(c) \gg \hbar \omega_D$, then correspondingly we have \( \tilde{V}_{\text{st}} \gg \hbar \omega_{\text{D}} \) and \( \Delta \gg \hbar \omega_{\text{D}} \). This is the dream that the ambient temperature superconductivity could be realized.
\begin{acknowledgments}
Project supported by the National Natural science Foundation of China (Grant No.10574163).
\end{acknowledgments}

\end{document}